# High-power magnetron transmitter as an RF source for superconducting linear accelerators

Grigory Kazakevich[*], Rolland Johnson, Gene Flanagan, Frank Marhauser,
Muons, Inc., Batavia, 60510 IL, USA

Vyacheslav Yakovlev, Brian Chase, Valeri Lebedev, Sergei Nagaitsev, Ralph Pasquinelli,
Nikolay Solyak, Kenneth Quinn, and Daniel Wolff,
Fermilab, Batavia, 60510 IL, USA

and Viatcheslav Pavlov,
Budker Institute of Nuclear Physics (BINP), Novosibirsk, 630090, Russia

A concept of a high-power magnetron transmitter utilizing the vector addition of signals of Continuous Wave (CW) magnetrons, injection-locked by phase-modulated signals, and intended to operate within a wideband control feedback loop in phase and amplitude, is presented. This transmitter is proposed to drive Superconducting RF (SRF) cavities for intensity-frontier GeV-scale proton/ion linacs, including linacs for Accelerator Driven System (ADS). The transmitter performance was verified in experiments with CW, S-Band, 1 kW magnetrons. A wideband dynamic control of magnetrons, required for the superconducting linacs, was realized using the magnetrons, injection-locked by the phase-modulated signals. The capabilities of the magnetrons injection-locked by the phase-modulated signals and adequateness for feeding of SRF cavities have been verified by measurements of the transfer function magnitude characteristics of single and 2-cascade magnetrons in the phase modulation domain, by measurements of the magnetrons phase performance and by measurements of spectra of the carrier frequency of the magnetrons. At the ratio of power of locking signal to output power of $\geq$ -13 dB (in 2-cascade scheme per magnetron, respectively) we demonstrated a phase modulation bandwidth of over 1.0 MHz for injection-locked CW single magnetrons and a 2-cascade setup, respectively. The carrier frequency spectrum (width of ~ 1 Hz at the level of <-60 dBc) of the magnetron, injection-locked by a phase-modulated signal, did not demonstrate broadening at wide range of magnitude and frequency of the phase modulation. The wideband dynamic control of output power of the transmitter model has been first experimentally demonstrated using two CW magnetrons, combined in power and injection-locked by the phase-modulated signals. The experiments with the injection-locked magnetrons adequately emulated the wideband dynamic control with a feedback control system, which will allow to suppress parasitic modulation of the accelerating field in the SRF cavities, resulted from mechanical noises, phase perturbations, caused by cavity beam loading and cavity dynamic tuning errors, low-frequency ripples of the magnetron power supplies, etc. The magnetron transmitter concept, tests of the transmitter models and injection-locking of magnetrons by phase-modulated signals are discussed in this work.

## I. INTRODUCTION

State of the art intensity frontier GeV-scale proton or ion superconducting linacs require CW RF sources to power SRF cavities, keeping the accelerating voltage phase and amplitude deviations to less than 1 degree and 1% of nominal, respectively. The average RF power to feed, for example, an ILC-type SRF cavity, providing an energy gain of ~20 MeV/cavity for a 1-10 mA average beam current, is a few tens to a few hundreds of kW.

The investment costs for an RF power system for large-scale projects (e.g. ADS facilities, etc.) are a significant fraction of the overall costs, if traditional RF sources as klystrons, Inductive Output Tubes (IOTs) or solid-state amplifiers are used. Utilization of MW-scale CW klystrons to power groups of the cavities can save costs to some extent, but in turn only allow a control of the vector sum of the accelerating voltage in the group. Accelerating voltage vector sum control has not been tested for driving SRF cavities for non-relativistic or weakly relativistic particles; it may be unacceptable for low-velocity particles since non-optimized values of phase and amplitude of the accelerating field in individual SRF cavities can cause emittance growth, [1], and may lead to beam losses.

The CW magnetrons based on commercial prototypes are more efficient and potentially less expensive than the above-mentioned RF sources, [2], thus utilization of the magnetron RF sources in the large-scale accelerator projects will provide significant reduction of capital and maintenance costs, especially since the CW magnetrons with power of tens to hundreds kW are well within current manufacturing capabilities.

---

[*]Current affiliation: Muons, Inc.,
552 North Batavia Ave., Batavia, IL 60510, USA
e-mail: gkazakevitch@yahoo.com; grigory@muonsinc.com

Mechanical oscillations of SRF cavities, including "microphonics" and oscillations caused by Lorentz-force, result in parasitic modulation of the accelerating field in the cavities. For ILC-type cavities, the modulation may have a bandwidth from tens to hundreds Hz. The modulation can be suppressed using electronic damping techniques based on a control of the feeding power, [3-5].

Additional perturbations of the accelerating field in SRF cavities are caused by dynamic tuning errors and beam loading. To suppress these perturbations, a phase control of the accelerating field is necessary.

Since the manufactured SRF cavities are not identical in mechanical and RF properties, these perturbations vary from cavity to cavity. Instantaneous phases and amplitudes of the mechanical oscillations are random in each cavity. Thus a simple concept to power multiple SRF cavities with phase-locked magnetrons, [6], is not applicable for the proton/ion superconducting accelerators. The concept most applicable for the accelerators is the powering of each SRF cavity by an individual vector controlled RF source providing a dynamic control of phase and of power.

A proof-of-principle of the phase control of a magnetron, injection-locked by a frequency (phase)-modulated signal first has been demonstrated modelling a transient process in the 2.5 MW, 2.8 GHz pulsed magnetron type MI-456A, forced (injection-locked) by a signal with varied frequency, [7-9]. The transient process model was verified with very good accuracy by the measurements of the magnetron frequency (phase) response, [ibid.]. Unlike the approach for vacuum tube LC generator developed by R. Adler, [10], assuming steady-state in oscillations injection-locked by an external signal, stable in frequency and phase, and then applied to magnetrons and described in numerous works, the techniques modelling the transient process allow computation of the magnetron frequency and phase response on the injection-locking signal, modulated in frequency (phase). I.e. this approach considers a dynamic control. Further analysis of the computed and measured response of the magnetron on the frequency and phase-modulated locking signal, [11], demonstrated an acceptable linearity and small phase errors in the response.

Measurements of the magnitude transfer characteristic and the phase performance of CW, S-band, 1 kW magnetrons, injection-locked by the phase-modulated signal, described in the presented work, demonstrated a wide bandwidth of the dynamic phase control. Performed measurements demonstrate that the injection-locking by the phase-modulated signal at the wide range of magnitude and frequency of the phase modulation does not broaden the very narrow magnetron phase noise spectrum. I.e. the magnetrons injection-locked by the phase-modulated signal are adequate RF sources for SRF cavities. The measurements and the presented analysis indicate that a wideband control of magnetrons by injection-locking phase-modulated signal will appropriately satisfy the requirements of intensity-frontier superconducting linacs.

Estimations and numerical modelling based on measurements of the transfer function demonstrates that power line related phase modulation sidebands of the injection-locked magnetrons associated with low-frequency phase pushing may be almost completely eliminated by closed loop feedback of the phase term in the Low Level RF (LLRF) controller. This closed loop also will suppress phase perturbations from cavity beam loading, cavity dynamic tuning errors and perturbations of the magnetron magnetic field induced by magnetron filament AC circuitry, which were observed in the presented work.

A dynamic power control of magnetrons injection-locked by the phase-modulated signal in the setup with power combining, first realized in this work, demonstrates the capability of the magnetrons for the wideband vector control of the accelerating field in SRF cavities.

The experimental tests, measurements, and numerical modelling performed with injection-locked magnetrons demonstrate proof-of-principle of the proposed magnetron transmitter concept to provide the dynamic vector control necessary to suppress all known low-frequency parasitic modulation in SRF cavities. The modelling and tests are discussed in this paper.

## II. A CONCEPT OF THE MAGNETRON TRANSMITTER CONTROLLED IN PHASE AND POWER

The proposed concept of high-power magnetron transmitter based on the injection-locked 2-cascade magnetrons, [12], is represented in Fig. 1.

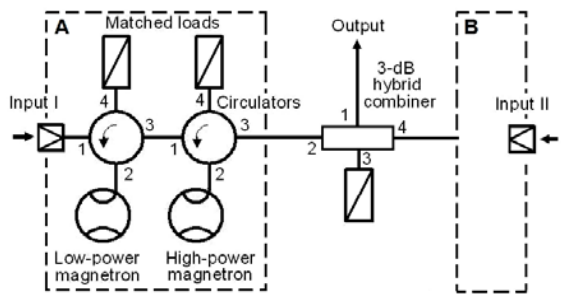

FIG 1. Block-diagram of the magnetron transmitter available for a wideband dynamic control in phase and power.

The transmitter consists of two identical channels (A and B) of cascaded injection-locked magnetrons with outputs combined by a 3-dB hybrid. For phase management the phases at inputs of both 2-cascade magnetrons are controlled simultaneously and equally, while the power management is provided by a control of phase difference at the inputs of the 2-cascade magnetrons. The 2-cascade injection-locked magnetrons composed from low-power and high-power magnetrons with series connection via circulators were proposed to use lower locking power (-35 to -25 dB, relative to the combined output power). This allows using hundreds Watts drivers for the transmitter with combined power of hundreds kW. It will decrease the capital cost of the solid-state drivers, injection-locking the 2-cascade magnetrons.

The transmitter scheme is acceptable for a wide dynamic power range. The transmitter efficiency is determined in general by the dynamic range of output power when the difference in powers of the 2-cascade magnetrons is insignificant, [13].

## III. TECHNIQUES TO STUDY AND TEST THE MAGNETRON TRANSMITTER CONCEPT

Proof-of-principle of the proposed CW magnetron transmitter based on the injection-locked tubes was demonstrated in experiments with S-band, CW, microwave oven magnetrons with output power of up to 1 kW. The magnetrons at most of described experiments operated in pulsed mode with pulse duration of a few ms at low repetition rate. This allowed employing lower average power RF and High Voltage (HV) components. Moreover, such an operation of the magnetrons allowed verifying the particular applicability of the transmitter concept for pulsed linacs accelerating long trains of bunches. Features of the transmitter in pulsed mode were experimentally studied and tested using two magnetrons that were chosen to be locked at the same frequency. The two magnetrons, tubes types 2M219J and OM75P(31), with a difference in the free run frequencies of about 5.7 MHz at an output power of ~500 W, were used in the experiments. The magnetrons have dissimilar Volt-Amps characteristics, but they were powered by a single pulsed modulator with partial discharge of a 200 μF storage capacitor, commutated by a HTS 101-80-FI IGBT (Behlke) fast switch. The modulator provides simultaneous operation of the two magnetrons with RF power up to 1 kW, each. A compensated divider has been used to power the magnetron with lower anode voltage when both magnetrons operated simultaneously. The level of ripple seen by the magnetron is negligible in the modulator, however, the capacitor introduces a voltage droop of approximately 0.4% to the 5 ms pulse when the modulator is loaded by the two 1 kW magnetrons. To protect the magnetrons and the modulator components from arcs the modulator has an interlock chain that rapidly interrupts the HV if the modulator load current exceeds 3.5 Amps. The HV modulator was powered by a HV switching power supply.

Table I summarizes operating parameters of the modulator.

Table I: Operating parameters of the modulator.

| Parameter | Symbol | Value |
|---|---|---|
| Output voltage | $U_{Out}$ | -(1-5) kV |
| Repetition rate | $f_{rep}$ | 0.25 Hz |
| Pulse duration | $t_{pulse}$ | 2.5 - 15 ms |
| Output current | $I_{Out}$ | 0.3-1.0 A |
| Short Circuit Interlock threshold | $I_{threshold}$ | 3.5 A |

The proposed transmitter features have been studied and tested using two modules with the magnetrons operating at the locking frequency of 2.469 GHz. Scheme of the magnetron module is depicted in Fig. 2.

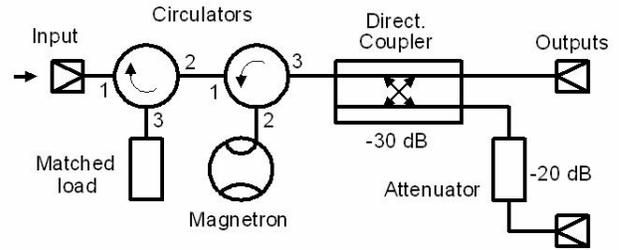

FIG. 2. The magnetron experimental module in which a CW magnetron operates as an injection-locked (forced) oscillator.

Each of the magnetrons was mounted on a WR430 waveguide, coupled with a waveguide-coax adapter. The waveguide section and the adapter were numerically optimized in 3D by CST Microwave Studio to minimize the RF reflection ($S_{11}$) at the operating frequency and thus to maximize the transmission ($S_{21}$), that resulted in $S_{11}$ = -26.3 dB and $S_{21}$ = -0.1 dB, respectively.

It has been verified that each magnetron can be injection-locked (at the same frequency) when operated in pulsed mode, pre-excited by a CW TWT amplifier driven by a CW signal generator (N5181A Agilent synthesizer) as it is shown in Fig. 3, [11].

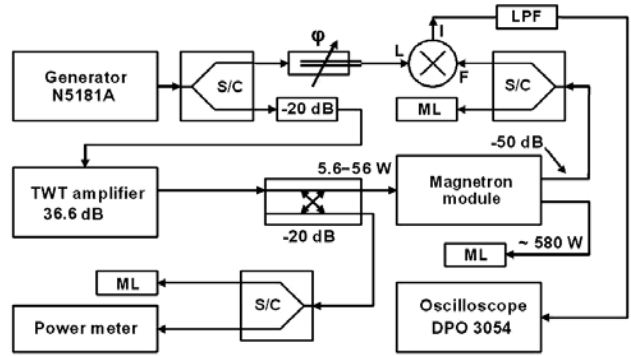

FIG. 3. Experimental setup with the phase detector to measure phase variations of the injection-locked magnetron. S/C is a 3-dB splitter/combiner; ML is a matched load.

A calibrated directional coupler, 3-dB splitter/combiner (S/C) and power meter were used to measure the locking power, $P_{Lock}$, while the magnetron output power, $P_{Out}$, was measured by the E4445A spectrum analyser which is not shown in this setup. Measured levels of the locking and output magnetron power are denoted in Fig. 3.

The setup was also used to measure intrapulse phase variations of the injection-locked magnetron relative to the synthesizer signal, which was used as a reference. The measurements were performed utilizing a simple phase detector, composed of a trombone-like phase shifter, φ, a double balanced mixer, and a Low Pass Filter (LPF).

Measured phase variation of the injection-locked magnetron at modulator pulse of ≈5 ms is shown in Fig. 4.

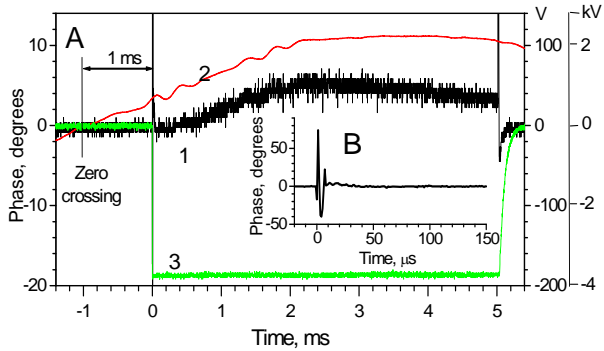

FIG. 4. A- phase variations of the injection-locked magnetron type 2M219J operating at pulse duration of 5 ms at $P_{Out}/P_{Lock}$ =9.6 dB, trace 1, left vertical scale. Traces 2 and 3 show shapes of the AC line voltage, and the magnetron HV pulse, right vertical scales, 100 V/div. and 2 kV/div., respectively. Inset B shows zoomed in time, the phase variation during first 150 μs.

The curve 1, Fig. 4A, shows phase variations during the 5 ms pulse. Inset 4B shows the phase variations during the first 0.15 ms. Both the traces show the variations composed from a low frequency (less than few kHz) and a high-frequency (more than few tens kHz) components, respectively.

The low-frequency components are regular and stable from pulse-to-pulse. They are phase perturbations resulting from transient processes in the injection-locked magnetron. The phase perturbations are caused in general by variation of the magnetron emission current and magnetron temperature at pulsed operation, alternative magnetic field induced by the magnetron filament circuitry (as it is shown below) and variation of the magnetron current associated with the discharge of the modulator storage capacitor. Note that the measurements of phase variations of the TWT amplifier driving the magnetrons do not indicate notable the low-frequency components.

The high-frequency components are stochastic noise.

Measured at the output of the injection-locked CW magnetron, the high-frequency phase noise magnitude of < 0.4 degrees (rms) in main part of the pulse (t ≥50 μs) reflects the contribution of the CW TWT wide-band amplifier. The performed measurements indicate that the stochastic noise of the magnetron itself is insignificant.

Note that the pulsed operation of the magnetrons with limited pulse duration and time of sampling (≤10 ms in these experiments) does not allow accurate measurements of spectral density of the low-frequency phase perturbations. Accurate measurements of the phase noise spectra of an S-band, 1 kW injection-locked magnetron, performed in CW mode, are described below.

Notable phase variation at the leading edge of the modulator pulse during of ~ 10 μs, Fig. 4B, at the rate of HV pulse rise of ≈5 kV/μs may result from phase pushing caused by several phenomena: multipactoring in the magnetron cavity when the pulsed high voltage is applied or/and variation of emitting properties of the magnetron cathode caused by cleaning of the emitting surface by back-streaming electrons. The time scales for both processes match the measured phase variation time. Operation of an SRF cavity would not suffer from the measured phase variation due to large filling time (a few ms) of the cavity.

The shape of the slow phase perturbations during the long pulse, trace 1, can be explained by phase pushing, resulting from competing processes: an increase of the magnetron current, most likely, because of overheating of the cathode surface caused by bombardment with back-streaming electrons and a decrease of the magnetron current associated with a droop of the magnetron voltage resulted from discharge of the modulator storage capacitor. The latter process dominates in the second half of the pulse. The thermal distortion of the magnetron geometry at pulsed operation and a magnetic field, induced by the magnetron filament AC circuitry may also contribute to the phase variations.

The trace 1 presented in Fig. 4 show phase variations magnitude of Δφ ≤5.3 degrees (peak-to-peak at t ≥ 50 μs) at pulse duration of ≈5 ms for the single magnetron operating in injection-locked mode. The magnitude of the phase variations depends on the locking power, Fig. 5.

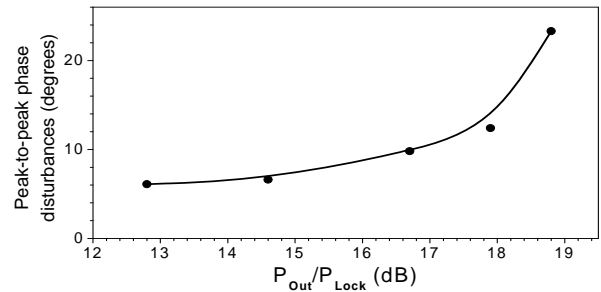

FIG. 5. Dependence of phase variation (peak-to-peak) of the injection-locked magnetron type 2M219J measured at the output power of 505 ± 5 W vs. the ratio of output power to locking power, $P_{Out}/P_{Lock}$.

Seen in Fig. 4, the synchronism between the distortions of the injection-locked magnetron phase variation, trace 1, and the distortions of the AC line voltage, trace 2, indicates an influence of a magnetic field induced by the magnetron filament circuitry on the magnetron phase stability. This was studied measuring phase variation (peak-to-peak) of the injection-locked magnetron vs. the time shift, $\Delta\tau$, between the moment of zero crossing of the magnetron filament current and triggering of the modulator.

The phase variations caused by the induced magnetic field were minimized, Fig. 6, by triggering the modulator with a 1 ms shift relative to the moment of zero crossing of the magnetron filament current, as shown in Fig. 4A.

Note that all experiments described in the presented work were conducted at nominal filament voltage of magnetrons.

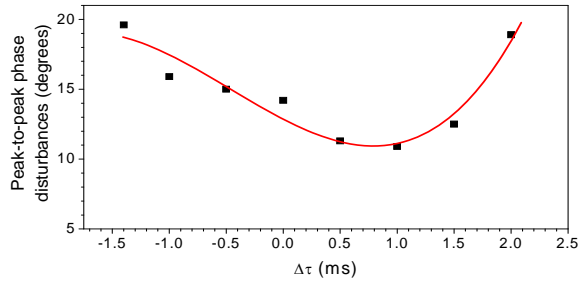

FIG. 6. Peak-to-peak phase variations of the injection-locked magnetron type OM75P(31) vs. the time shift $\Delta\tau$, measured at $P_{Out}/P_{Lock}$=15.8 dB. Solid curve shows a polynomial fit of the measurements.

The phase pushing value of ~1.5 deg/1% or ~ 500 deg/A, associated with the magnetron current variation was evaluated from the measurements at the ratio $P_{Out}/P_{Lock}$ ~16 dB.

## IV. EXPERIMENTAL VERIFICATION OF THE 2-CASCADE MAGNETRON CONCEPT

Operation of the 2-cascade injection-locked magnetron, in which the first injection-locked, lower-power magnetron is used for injection-locking of the second one with higher power, has been tested in the experimental setup, Fig. 7, [13], composed of the two magnetron modules connected in series via an attenuator. This provided injection-locking in the second magnetron by the attenuated signal from the first injection-locked magnetron.

Both of the injection-locked magnetrons were powered simultaneously by the modulator at the pulse duration of ≈ 5 ms. The first magnetron was injection-locked by the CW TWT amplifier, while the second magnetron was locked by the pulsed signal generated by the first injection-locked magnetron. The magnetron having lower anode voltage has been powered through the compensated divider. The output powers of the magnetrons denoted in Fig. 7 were measured by the E4445A spectrum analyser, which is not shown in this setup. Phase variations of the 2-cascade injection-locked magnetron were measured at various attenuator values in the range from 13 dB to 20 dB, [ibid.].

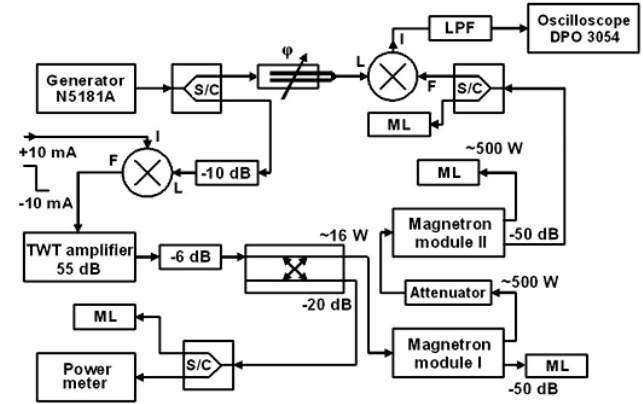

FIG. 7. Experimental setup to measure phase variation of the 2-cascade injection-locked magnetron.

Trace of the phase variation of the 2-cascade injection-locked magnetron setup measured at the ratio $P_{Out}/P_{Lock} \approx$ 30 dB, considering the attenuator value of 15 dB, is plotted in Fig. 8.

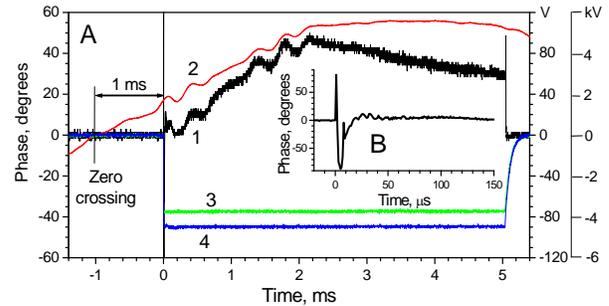

FIG. 8. A- phase variations of the 2-cascade injection-locked magnetron measured at pulse duration of ≈ 5 ms at the attenuator value of 15 dB, trace 1, left vertical scale. Trace 2 shows voltage in the AC line, right vertical scale, 40 V/div. Traces 3 and 4 show shapes of HV pulses applied to magnetron I and magnetron II, respectively, right scale, 2 kV/div. Inset B shows zoomed in time trace 1.

The measured phase variation of the 2-cascade injection-locked magnetron, trace 1, resembles in shape traces of the injection-locked single magnetrons. However, close location of the both magnetron modules in the used setup caused larger impact of leakage fields from both filament transformers. This results in phase variation about of 15-20 deg. (per magnetron) at the power of locking signal of -15 dB. Nevertheless, the plotted trace 1 clearly demonstrates that the 2-cascade magnetron was still operating in injection-locked mode. This demonstrates proof-of-principle of the 2-cascade injection-locked magnetron. Faster droop of the phase trace at t ≥2.5 ms is caused by the phase pushing resulting from enlarged discharge of the modulator capacitor loaded by two magnetrons.

Measured high-frequency noise magnitude at t ≥50 μs is ≤ 0.7 degrees (rms) at the output power of ≈ 500 W. The measured phase trace 1, Fig. 8, demonstrates operation of the experimental setup of the 2-cascade injection-locked magnetron at a ratio of the output power to the locking power of ≈ 30 dB, considering the attenuator value.

The phase response of the 2-cascade injection-locked magnetron on the 180 degrees phase flip has been evaluated using the setup shown in Fig. 7, [13]. The 180 degrees phase flip in the TWT drive signal is accomplished with a pulse generator and double balanced mixer on the TWT amplifier input. The measured transient process of the injection-locked magnetron response, Fig. 9, on the 180 degrees phase flip takes ~ 300 ns at the flip duration of ≈ 15 ns. The time of the response is ~750 periods of the magnetron oscillation.

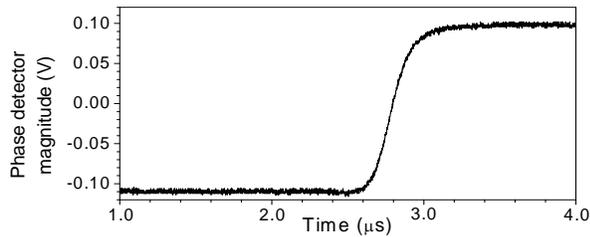

FIG. 9. Response of the injection-locked 2-cascade magnetron on the 180 degrees phase flip measured at ratio of the output power to locking power ≈27 dB; the phase detector calibration is ≈ 0.8 degrees/mV.

Plot shown in Fig. 9 indicates quite wide bandwidth in the phase response of the 2-cascade magnetrons. More accurate measurements of the bandwidth of the phase control of the injection-locked magnetrons were performed using the phase modulation method described below.

## V. EXPERIMENTAL VERIFICATION OF THE POWER CONTROL IN THE MAGNETRON TRANSMITTER

A setup to study the power control of the injection-locked CW magnetrons with 3-dB 180 degrees hybrid combiner in static and dynamic regimes is shown in Fig. 10.

A phase shifter, (trombone) $\varphi_{II}$, was used to vary the power combined at port "Σ" of the 3-dB 180 degrees hybrid by variation of the phase difference in RF signals injection-locking the magnetrons in a static regime. The analogue phase shifter JSPHS-2484 controlled by voltage was added for pulsed power control in a dynamic regime. The spectrum analyser E4445A was used to measure in static regime the combined power and individual power of each magnetron. The phase detector with the phase shifter $\varphi_I$, double balanced mixer and LPF was used to measure phase deviations in the power combining scheme in static regime and a calibrated diode detector was used to measure power of the combined signal in dynamic regime.

The measured ratios of the output power to the locking power of the magnetrons, shown in Fig. 10, were 17.6 dB and 14.9 dB, respectively.

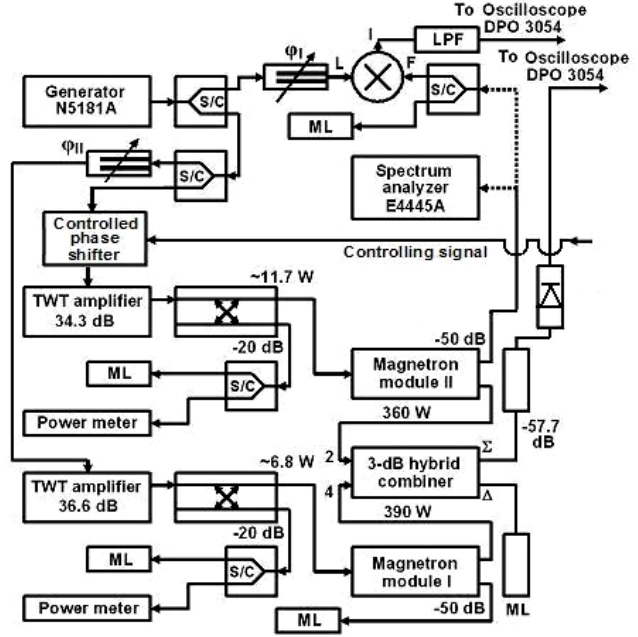

FIG. 10. Setup with the injection-locked magnetrons for test of the power control by power combining.

Phase variations of the injection-locked magnetrons with power combining during 5-ms pulse are shown in Fig. 11.

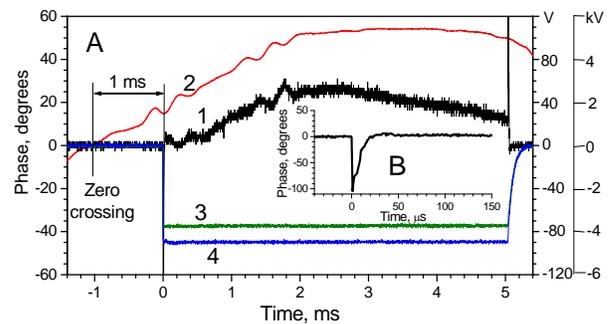

FIG. 11. A- phase variations at the output "Σ" of the hybrid, trace 1, left vertical scale. Trace 2 shows the AC line voltage, right vertical scale, 40 V/div. Traces 3 and 4 shows pulsed voltages feeding the magnetrons I and II, respectively, right vertical scale, 2 kV/div. Inset B shows zoomed in time trace 1.

At the static measurements the trombone $\varphi_{II}$ length, Fig. 10, has been chosen to provide maximum signal at the hybrid port "Σ", the phase shifter JSPHS-2484 control was OFF.

Part of the trace 1 at t ≥ 50 μs, Fig. 11, measured with the phase detector, has a smooth shape with high-frequency phase noise magnitude of ≤ 0.5 degrees (rms). The phase trace resembles the traces of single magnetrons or 2-cascade magnetron. The phase variation magnitude roughly corresponds to the magnitude of single magnetron (15-20 deg.), measured when the magnetrons are located close as it was in the used setup. Larger phase variation at t ≥ 2.5 ms for magnetrons with power combining (similarly to phase variation of the 2-cascade magnetron) in comparison with a single magnetron results from larger phase pushing in the injection-locked magnetrons because of larger discharge of the modulator storage capacitor loaded by the two magnetrons.

The vector power control in magnetrons with power combining vs. the phase difference, resulted from variation of the phase shift by the trombone $\varphi_{II}$ or by the controlled calibrated phase shifter, measured in static and dynamic regimes, respectively, is shown in Fig. 12.

The dynamic control was realized by phase modulation of the signal, injection-locking the magnetron II, Fig. 10. The modulation was performed by a sequence of pulses with period of ≈30 μs controlling the phase shifter JSPHS-2484. At the dynamic control the combined power was measured by the calibrated diode detector; the modulated phase shift was measured by the phase detector.

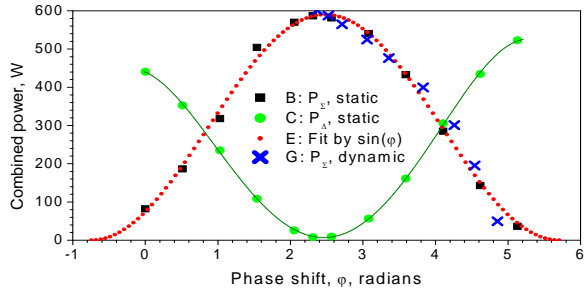

FIG. 12. Control of combined power of the injection-locked magnetrons by the phase difference. Dots B and C present power variation at the combiner ports "Σ" and "Δ", respectively, measured in static regime. Dots G show power variation at the combiner port "Σ" measured in dynamic regime. Dots E show fit of the plots B and G by sin(φ) function.

The plots show measured power at the combiner outputs "Σ",dots B and G, and "Δ", dots C, considering the hybrid insertion losses of 0.4 dB and 0.7 dB, respectively. Plot E shows fit of the plots B and G by function sin(φ). Note that the simple phase detector does not provide a sufficient accuracy at the phase shift > 1.5 rad.

Agreement of the measured combined power, plots B and G with the fit trace, plot E, verifies that the utilized method of power control does not disturb injection-locking of the magnetrons. The plots B, G and E indicate an acceptable linearity and low phase errors in response of the injection-locked 1 kW, CW magnetrons at the dynamic phase control as it was observed earlier in operation of the single 2.5 MW pulsed magnetron locked by frequency (phase)-modulated signal.

Real time performance of the dynamic power control of the injection-locked magnetrons with power combining, realized by the phase modulation with the analogue phase shifter JSPHS-2484, controlled by the rectangular pulses of voltage, is shown in Fig. 13.

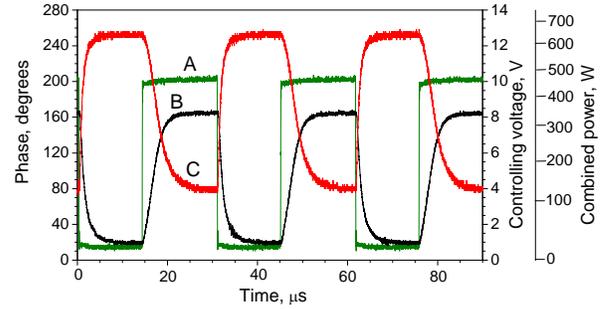

FIG. 13. Dynamic power control of the injection-locked magnetrons with power combining. Trace A shows shape of signals controlling the phase shifter, first right scale. Trace B is the phase variations at the output of the magnetron II measured by the phase detector, left scale. Trace C shows power measured at port "Σ" of the hybrid combiner at the phase shifter control, second right scale.

The measurements performed at pulse duration of ≈5 ms do not indicate notable distortions in the power modulation measured at the combiner port "Σ" along the pulse. At the measurements the phase shifter $\varphi_{II}$ was tuned to get maximum power at the port "Σ" when the signal controlling the phase shifter JSPHS-2484 was OFF. The phase detector trombone $\varphi_I$ was tuned to avoid saturation at measurements of the phase modulation of the magnetron in the module II.

Note the bandwidth of the analogue phase shifter is ≈50 kHz. The bandwidth of the power control of the combined magnetrons shown in plots B and C is limited by this value, but not by the bandwidth of the magnetron phase control, which is much wider as it is shown below.

## VI. FEATURES OF MAGNETRONS, INJECTION-LOCKED BY A PHASE-MODULATED SIGNAL

Features of the dynamic control of the injection-locked magnetrons were analyzed basing on the earlier results of modeling the transient processes in the 2.8 GHz, 2.5 MW magnetron, [7-9], and study of the transfer characteristic, phase performances, and phase noise spectra measured at various setups with CW, 1 kW magnetrons, injection-locked by the phase-modulated signal.

Obtained in the earlier works, the abridged equation describing transient processes caused by the phase pushing and modulation of the locking signal in the injection-locked magnetron, decoupled from the load, can be written as:

$$\left\{\frac{d}{dt}+\frac{\omega_{0M}}{2Q_{LM}}(1-i\varepsilon_M)\right\}\widetilde{V}_M = \frac{\omega_{0M}}{Q_{EM}}\widetilde{V}_{FM} - \frac{\omega_{0M}}{2Q_{EM}Y_{0M}}\widetilde{I}_M. \quad (1)$$

Here: $Q_{LM}$, and $Q_{EM}$ are loaded and external magnetron Q-factors, respectively, $\omega_{0M}$ is the eigenfrequency of the magnetron cavity, $\varepsilon_M = \tan\psi \approx 2Q_{LM}(\omega_{0M}-\omega)/\omega_{0M}$ is the detuning parameter, $\omega$ is the frequency (time-dependent in common case) of the locking signal, $\widetilde{V}_M$ and $\widetilde{V}_{FM}$ are complex amplitudes of the oscillation in the magnetron cavity and in the wave locking the magnetron, respectively, $Y_{0M}$ [1/Ohm] is the external waveguide conductance of the magnetron cavity, $\widetilde{I}_M$ is the complex amplitude of the first harmonic magnetron current. Terms $\frac{\omega_{0M}}{Q_{EM}}\cdot\widetilde{V}_{FM}$ and $\frac{\omega_{0M}}{2Q_{EM}\cdot Y_{0M}}\cdot\widetilde{I}_M$ describe modulation of the locking signal and phase pushing, respectively, $\psi$ is the angle between sum of the phasors $\widetilde{V}_{FM}$ and $\widetilde{I}_M$ and the phasor $\widetilde{V}_M$, taken with corresponding coefficients, [14].

The equation (1) has been solved numerically at measured and computed time-dependent magnitudes of the phasors $\widetilde{V}_M$, $\widetilde{V}_{FM}$ and $\widetilde{I}_M$.

Computed and measured responses of the 2.5 MW magnetron on the frequency (phase)-modulated locking signal are presented in Fig. 14, [11], plot G and D, respectively. Plot B shows time-dependent variation of the frequency of the signal at a power of -18 dB, locking the magnetron.

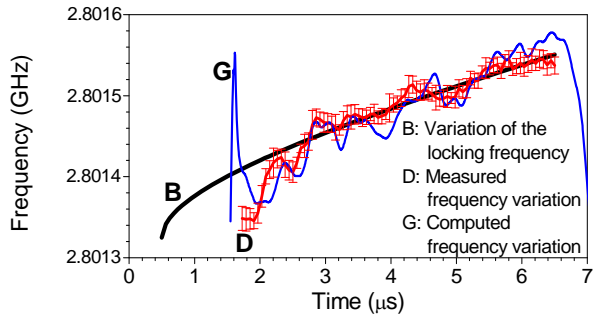

FIG. 14. Variation of the computed (G), and measured frequency (D), of the 2.5 MW, 2.8 GHz pulsed magnetron vs. variation of the locking frequency (B) in time domain.

The measurements show that time-to-lock of the magnetron does not exceed 2 μs, including the time to establish magnitude of the locking signal which is ~ 0.5 μs. After the time-to-lock (t ≥2.5 μs) the magnetron well repeats frequency variations of the forcing signal, i.e. it is locked by the frequency (phase)-modulated signal.

Note the ripple in the computed trace G at t ≥ 2.5 μs (when the magnetron was locked by the frequency-modulated signal) is produced in general from the data acquisition system measuring the magnetron voltage, current, amplitudes in forward and reflected waves and partially from limited statistics in simulation of the locking signal. The data acquisition system was not isolated well enough from the 5 MW modulator grounding bus-bar, [9]. This caused noise in measurements and produced "ripple" in the computation. Regardless, the trace of the measurements of the magnetron frequency, plot D, at t ≥ 2.5 μs is in good coincidence with plot of the frequency-modulated locking signal, trace B.

The injection-locked magnetron rms phase error was computed from the measured frequency variation, plot D, by integration over the filling time of the source of the locking signal, [11]. The obtained value of the rms phase error does not exceed 0.4 degrees. The measured plot D in Fig. 14 demonstrates an acceptable linearity of the frequency (phase) response of the magnetron injection-locked by the frequency (phase)-modulated signal.

A dynamic phase control of the 1 kW, CW magnetrons, considered as a transient process, was realized in setups shown in Figs. 3, 7, 10, using a harmonic phase modulation of the injection-locking signal (in the synthesizer) at the frequency $f_{PM}$ and at the magnitude of the modulation of 20 degrees. The harmonic modulation improves accuracy in determination of the magnetrons phase performance vs. $f_{PM}$ at the transient process, caused by the dynamic control.

The magnetron phase performance at the dynamic phase control was determined as the angle of rotation of phasor of voltage in the wave at the magnetron output vs. $f_{PM}$. According to [8, 9], the phasor of voltage at the magnetron output, $\widetilde{V}_{MO}$, is expressed as: $\widetilde{V}_{MO} = \widetilde{V}_M - \widetilde{V}_{FM}$. At $|\widetilde{V}_M| \gg |\widetilde{V}_{FM}|$, $\widetilde{V}_{MO} \approx \widetilde{V}_M$. Behaviour of the phasor $\widetilde{V}_M$ vs. phase modulation of the locking signal $\widetilde{V}_{FM}$ can be evaluated considering equation (1) in a steady-state. In this case the equation (1) is transformed into equation (2), [14].

$$\widetilde{V}_M = \cos\psi\cdot\exp(i\psi)\cdot\left(\frac{2Q_{LM}}{Q_{EM}}\widetilde{V}_{FM} - \frac{Q_{LM}}{Q_{EM}}\widetilde{I}_M\right). \quad (2)$$

From this equation follows that variation of phase of the phasor $\widetilde{V}_{FM}$ varies phase of the phasor $\widetilde{V}_M \approx \widetilde{V}_{OM}$. It allows to find angle $\theta$ of the phasor $\widetilde{V}_{OM}$ rotation vs. $f_{PM}$ (at constant magnetron current) measuring difference of normalized phasors $\widetilde{V}_{OM}$ and $\widetilde{V}_{FM}$ by the phase detector.

The angle $\theta$ is determined as: $\theta \cong a\cos(1-V_O/V_{PM})$, or $\theta \cong a\sin(V_O/V_{PM}-1)+\pi/2$ at $V_O \leq V_{PM}$ and $V_O > V_{PM}$, respectively. Here: $V_O$ is voltage measured at output of the phase detector, $V_{PM}$ is voltage at output of the phase detector corresponding magnitude of the phase modulation. The phasor $\widetilde{V}_{OM}$ rotation angle vs. frequency of phase modulation, $f_{PM}$, for various setups, various values of locking power, and for magnitude of the phase modulation of 20 degrees, is plotted in Fig. 15. Inaccuracy of the plots does not exceed 20%.

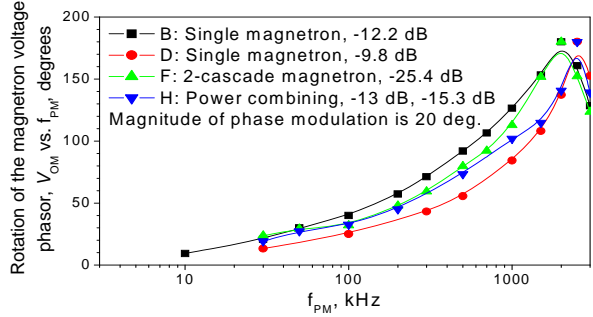

FIG. 15. Angle of rotation of the phasor of voltage in the wave at output of the injection-locked magnetrons resulted from the transient process, and measured in various setups at various locking power vs. the modulating frequency, $f_{PM}$.

Shown in Fig. 15 plots demonstrate that the phase performance of the magnetron injection-locked by phase-modulated signal depends on power of the locking signal.

Note that the angle of the phasor rotation characterises the phase modulation index, but does not represent a pole on the magnetron phase transfer characteristic. Even the angle a few radians does not disturb injection locking in the magnetrons as it is shown in Fig. 15 at $f_{PM}$ ~2 MHz. The magnetrons are still keeping the same carrier frequency.

The group delay of magnetron response on the phase-modulated signal computed from data shown in Fig. 15 depends on the locking power and for all tested setups does not exceed ~40 ns, [15]. The value has a simple physical meaning: the group delay for forced oscillation at carrier frequency, $f_M$, of magnetron is determined by the filling time of the magnetron cavity, which is proportional to $Q_{LM}/f_M$. I.e. the transient process of the phase control in the injection-locked magnetrons is averaged over the magnetron filling time. Thus the magnetron is still injection-locked at the carrier frequency $f_M$, if the phase modulation magnitude is $<2\pi$ during the filling time.

Measured carrier frequency spectra of a magnetron injection-locked, by phase-modulated signal are shown in Fig. 16. The measurements at the bandwidth resolution of 1.0 Hz were performed by the Agilent MXA N9020A Signal Analyzer using the microwave oven, S-band, 1 kW magnetron, operating in CW mode at $P_{Out}/P_{Lock}$= 12.7 dB, output power of ~ 850 W, carrier frequency of the locking signal of 2.451502 GHz. Setup of the measurements looks like shown in Fig. 2. Loss of power at the carrier frequency, Fig. 16, trace B, at large angle of the phase modulation results from redistribution of the magnetron power into sidebands at large index of the modulation. The sidebands differing from the carrier frequency by 11.3 and 15 Hz result from the locking system.

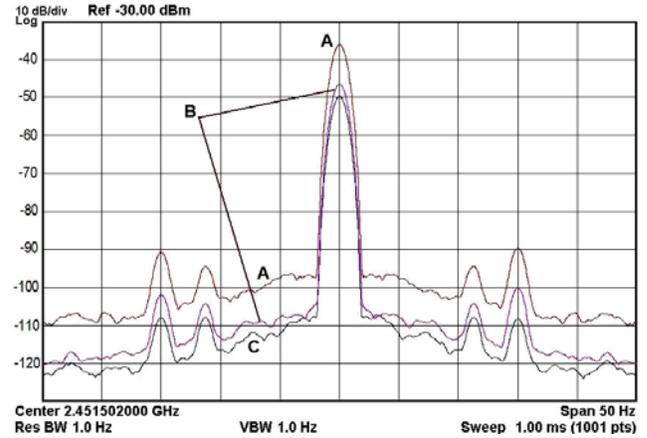

Fig. 16. Carrier frequency spectra of the magnetron injection-locked by a signal without, trace A, and with the phase-modulation. Trace B shows the spectrum of the magnetron injection-locked by the phase-modulated signal at magnitude and frequency of the modulation of 3 radians and 2 MHz, respectively. Trace C shows the carrier frequency spectrum of the locking system (N5181A generator and TWT amplifier). Scales in vertical and horizontal are: 10 dB/div and 5 Hz/div, respectively.

No broadening of the carrier frequency spectra was observed in the range of modulating frequency of (0-2 MHz) and large magnitude (up to few radians) of the phase modulation. Plotted in Fig. 16 the magnetron carrier frequency spectrum at level of -60 dBc is 3.8 Hz, while the carrier frequency spectrum of the locking signal is 3.68 Hz at level of -60 dBc. This indicates that the own width of the carrier frequency spectrum for the injection-locked magnetron, measured at the level <-60 dBc at the bandwidth resolution of 1.0 Hz, can be evaluated as ~1 Hz and the phase noise of the magnetron injection-locked by phase-modulated signal is very low. Hence, the bandwidth of the carrier frequency of the magnetron, injection-locked by the phase-modulated signal, is adequate to bandwidth of SRF cavities of superconducting linacs.

The bandwidth of the phase management of CW injection-locked magnetrons necessary for modelling of the control loop has been determined by measurements of the magnitude transfer characteristic of the CW, S-band magnetrons. The measurements, Fig. 17, were performed at pulsed operation of magnetrons in setups shown in Figs. 3 and 7.

At the measurements was used phase modulation with magnitude of 0.07 rad. in the synthesizer. The transfer function magnitude characteristics were measured by the Agilent MXA N9020A Signal Analyzer in the phase modulation domain for various ratios $P_{Out}/P_{Lock}$.

The plotted transfer magnitude characteristics (rms values) were averaged over 8 pulses for the injection-locked single 2M219J magnetron and the 2-cascade magnetron setup. Non-flatness of the synthesizer phase characteristic has been measured and taken into account.

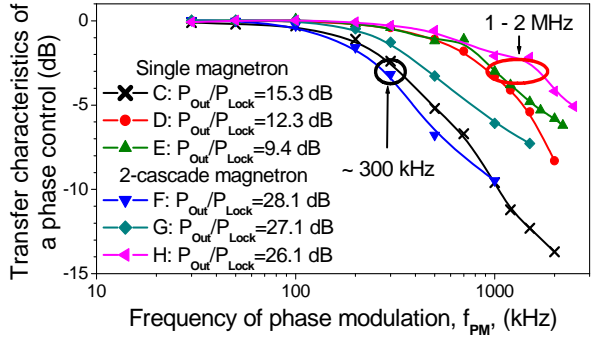

FIG. 17. Transfer function magnitude characteristics (rms values) of the phase control measured in phase modulation domain with single and 2-cascade injection-locked magnetrons at various ratios $P_{Out}/P_{Lock}$, at $P_{Out} \approx 450$ W.

The transfer characteristics of the magnetrons demonstrate wide bandwidth that allows for fast phase control of the magnetrons.

The cutoff frequency of the phase modulation controlling the injection-locked magnetrons depends, as it is seen in Fig. 17, on ratio of the magnetron output power to power of the locking signal. The measured cutoff frequency of the phase modulation is $\approx 300$ kHz at the locking power relative to the output power (per magnetron in the 2-cascade scheme) at about -14 dB or less, while at the locking power relative to the output power (per magnetron in the 2-cascade scheme) at more than about -13 dB, the cutoff frequency of the phase modulation is $\geq 1.0$ MHz.

The transfer characteristics measured in the phase modulation domain, Fig. 17, implies that a Low Level RF controller may have a closed loop with a bandwidth of $\geq 100$ kHz and will be able to suppress all expected system low-frequency perturbations, such as parasitic low frequency phase modulation (hundreds of Hz) and phase perturbations from SRF cavity beam loading, the cavity dynamic tuning errors, the low-frequency perturbations caused by magnetron power supplies ripples, etc.

For example, the parasitic modulation caused by HV power supply ripple at frequency $f_r=120$ Hz will be suppressed within a phase feedback loop with integral gain $I=1.2 \cdot 10^7$ rad./s, by $\approx 20 \cdot \log(I/2\pi \cdot f_r) \approx 84$ dB.

Influence of the high-frequency phase noise of injection-locked magnetrons on the accelerating field in the SRF cavity has been numerically simulated with a simple model of a proportional-integral (PI) feedback phase loop around a superconducting cavity with a broad-band noise, Fig. 18.

Shown in this model a 200 Hz half bandwidth low-pass filter models the cavity base-band response, a 400 kHz bandwidth noise source represents the phase noise of the magnetron and a 2 μs delay represents all system group delay. The PI loop is setup with a proportional gain of 200 and integral gain $1.2 \cdot 10^7$ rad./s.

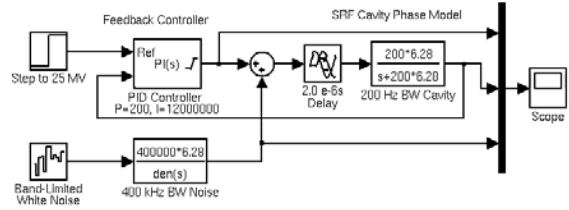

FIG. 18. Model of a LLRF system controlling accelerating field in SRF cavity. The loop proportional gain, P, is 200, the integral gain, I, is $1.2 \cdot 10^7$ rad./s, the group delay is 2 μs.

The numerical modelling, of the scheme shown in Fig. 18, demonstrate that the broad band noise associated with the greatly exaggerated magnetron high-frequency noise is suppressed by the controller with the PI loop including the SRF cavity by $\approx 50$ dB for peak-to-peak measurements, Fig. 19.

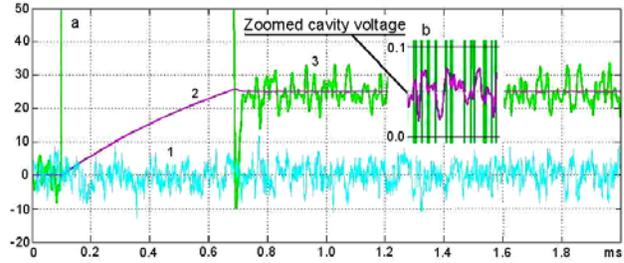

FIG. 19. a: curve 1 is the 400 kHz bandwidth disturbance, curve 2 is cavity voltage, curve 3 is RF drive. Vertical scale is 10 MV/division. The inset b presents zoomed in $\approx 300$ times (in vertical) trace of the cavity voltage, curve 2, in time domain. Vertical scale in the inset "b" is 0.1 MV/division.

The measurements and simulation indicate that the RF source based on magnetrons, injection-locked by the phase modulated signal, dynamically controlled by a wideband feedback system, can provide stability of phase and amplitude of the accelerating field required for SRF cavity.

Performed measurements utilizing the method of wideband phase modulation of the injection-locking signal, including measurements of the magnitude transfer characteristics, the magnetron performance at the dynamic phase control and the phase noise spectra adequately emulate wideband feedback system of the dynamic control. Results of the measurements substantiate the utilization of magnetrons injection-locked by phase-modulated signal to feed the SRF cavities of intensity-frontier linacs. This is reflected in the proposed transmitter concept. Based on the measurements and simulations we expect to satisfy the requirements of the superconducting linacs in stability of phase and amplitude of the accelerating field with the proposed magnetron transmitter.

Note that experiments described in [2, 16, 17] first verified operation of the injection-locked magnetron within a closed feedback loop.

## VII. SUMMARY

Presented are the results of a series of tests conducted on injection-locked S-band, 1 kW microwave oven magnetrons. The motivation for the R&D is verification of capabilities of magnetrons injection-locked by a phase-modulated signal allowing wideband dynamic amplitude and phase control in experimental model of the high-power transmitter. Injection-locked magnetrons are very low phase noise, efficient RF sources with a promising future for feeding SRF cavities for particle accelerators. While the proposed technique utilizes additional complexity, the cost savings of a successful system is significant in comparison to alternatives such as klystrons, IOTs, or solid-state amplifiers. Features of dynamic control of the magnetrons injection-locked by phase-modulated signal were studied with the 1 kW CW tubes operating in general in pulsed mode at various setups, modelling experimentally all active components of the proposed RF source based on magnetrons. The features are similar to these observed earlier in operation of 2.5 MW pulsed magnetron locked by frequency (phase)-modulated signal. What was shown is that injection locking of cascaded magnetrons looks to be feasible, this being a necessary requirement for RF powers exceeding 10 kWatts due to the low "locking gain" of 12-15 dB per magnetron. The dynamic power control in the described above injection-locked setup has been verified experimentally. A wideband dynamic control by phase modulation of the injection-locking signal has been verified for all setups, modelling operation of the proposed magnetron transmitter. Measured in CW mode the carrier frequency spectrum of the magnetron injection-locked by the wideband phase-modulated signal is ~1 Hz at the level < -60 dBc. The carrier frequency bandwidth is adequate to bandwidth of SRF cavities of intensity-frontier linacs. The measured phase modulation bandwidth of over 1.0 MHz appears to be adequate for the wideband dynamic phase and power control. The low-frequency phase perturbations less than 45 degrees measured in tests of active components of the transmitter can be suppressed by the phase control system. High-frequency phase noise of less than 1 degree (rms) measured in the tests of the proposed magnetron source is suitable for the intensity-frontier linacs requirements. Techniques using injection-locking phase-modulated signal adequately emulate wideband feedback system of a dynamic control. Finally, the proof of principle of the proposed transmitter concept has been demonstrated.

## ACKNOWLEDGMENT

This work has been supported by the US DOE grant DE-SC0006261 and collaboration Muons, Inc. – Fermilab. We thank Dr. Yu. Eidelman for useful discussions.